# Viral spread with or without emotions in online community

Andrzej Jarynowski[1,2], Jarosław Jankowski[3], Anita Zbieg[4,5]

[1]Jagiellonian University, Institute of Physics, Kraków, Poland
[2]Stockholm University, Department of Sociology, Stockholm, Sweden
[3]West Pomeranian University of Technology, Szczecin, Poland
[4]Wrocław University of Economics, Wrocław, Poland
[5]University of Wrocław, Institute of Psychology, Wrocław, Poland
andrzej.jarynowski@sociology.su.se, jjankowski@wi.zut.edu.pl, anita.zbieg@gmail.com

**Abstract.** Diffusion of information and viral content, social contagion and influence are still topics of broad evaluation. We have studied the information epidemic in a social networking platform in order compare different campaign setups. The goal of this work is to present the new knowledge obtained from studying two artificial (experimental) and one natural (where people act emotionally) viral spread that took place in a closed virtual world. We propose an approach to modeling the behavior of online community exposed on external impulses as an epidemic process. The presented results base on online multilayer system observation, and show characteristic difference between setups, moreover, some important aspects of branching processes are presented. We run experiments, where we introduced viral to system and agents were able to propagate it. There were two modes of experiment: with or without award. Dynamic of spreading both of virals were described by epidemiological model and diffusion. Results of experiments were compared with real propagation process - spontaneous organization against ACTA. During general-national protest against new antypiracy multinational agreement - ACTA, criticized for its adverse effect on e.g. freedom of expression and privacy of communication, members of chosen community could send a viral such as Stop-ACTA transparent. In this scenario, we are able to capture behavior of society, when real emotions play a role, and compare results with artificiality conditioned experiments. Moreover, we could measure effect of emotions in viral propagation. As theory explaining the role of emotions in spreading behaviour as an factor of message targeting and individuals spread emotional-oriented content in a more carefully and more influential way, the experiments show that probabilities of secondary infections are four times bigger if emotions play a role.

# 1 Introduction

The studies that direct attention to disease outbreaks [1], diffusion of innovation process [2], social influence mechanism [3], social contagion or cascades of influence patterns [4] usually based on similar phenomenon: an epidemic spread (e.g. pathogen, information content, opinions, behaviors, emotions) within a network of social relations. Such epidemics may follow variety of models, depending on the spreading properties, content, incentives, risk, individual attitudes of sender and recipient to the content and other factors [5]. Due to the fact that the knowledge of such a spread specifics is essential for the epidemiologist or sociologist researchers are constantly evaluating models and real-world data [6]. Researches in the field of online communities, virtual worlds and massively multiplayer platforms and games, among other relates to users engagement and social dynamics allow to study designed experimental setups not possible in offline communities [7]. Nevertheless, the empirical network studies seem to be promising, relatively little has been done in this area. Recently, the knowledge of complex system tools for sociology and medicine such as networks has undergone an accelerating growth, however all models of such system are incomplete with real data, especially register-based [8]. Network theory is useful when it comes to study nature from a systems perspective, and there are several examples in which it has helped understanding the behavior of complex systems [9]. Social interactions are good example in which a network perspective is important to understand systems behavior [5]. However, viral spread and even society itself change in time at different scales and in different ways. Dynamics from this perspective have not been studied in detail and an integrative framework is missing [10], especially for such an event like Stop-ACTA[1] movement formation. The research presented in this paper is targeted to online social platforms with the ability to capture different forms of users' behaviors: communications, activity and transfers among users [11]. The main motivation in the current research is to observe human action systems more by analyzing their behaviors related directly to viral campaign and stages of the participation for the viral action.

# 2 Experimental setup and background framework

During the research, there were data from two artificial (sending an avatar gesture) and one natural (sending an anti-ACTA transparent or mask) viral actions with different characteristics from social platforms working in a form of virtual world situated in one of Polish internet virtual world. Periodicity of users' activity affects our studies [Fig. 1]. For that reason real time pictures of their activity can be applied only on short time scale (up to one day). On the other hand, generalized, not time, but sequence oriented series analyses give us opportunity to look into system, which elements (users) were replaced by new ones. In this virtual world, users can wear avatar [Pic. 1 right] and that items are core of our study. In interesting as cases user can get access to specific element of avatar only by invitation, which he/she can spread with-

1 The Anti-Counterfeiting Trade Agreement (**ACTA**), a treaty signed in October 2011 to establish international standards for intellectual property rights enforcement.

out any limitations to other active users. In all actions, users were spreading virtual goods like avatars using viral mechanism to their friends. The first viral action was based on sending gifts to friends and the senders' motivation to spread those gifts was not incentivized. The second was based on incentives and competition among users to spread visual elements of avatars among their friends. The third – natural study was performed when Stop-ACTA movement[2] was starting in Poland. Stop-ACTA stand up [Pic. 1 left], which waved emotional value, could be spread.

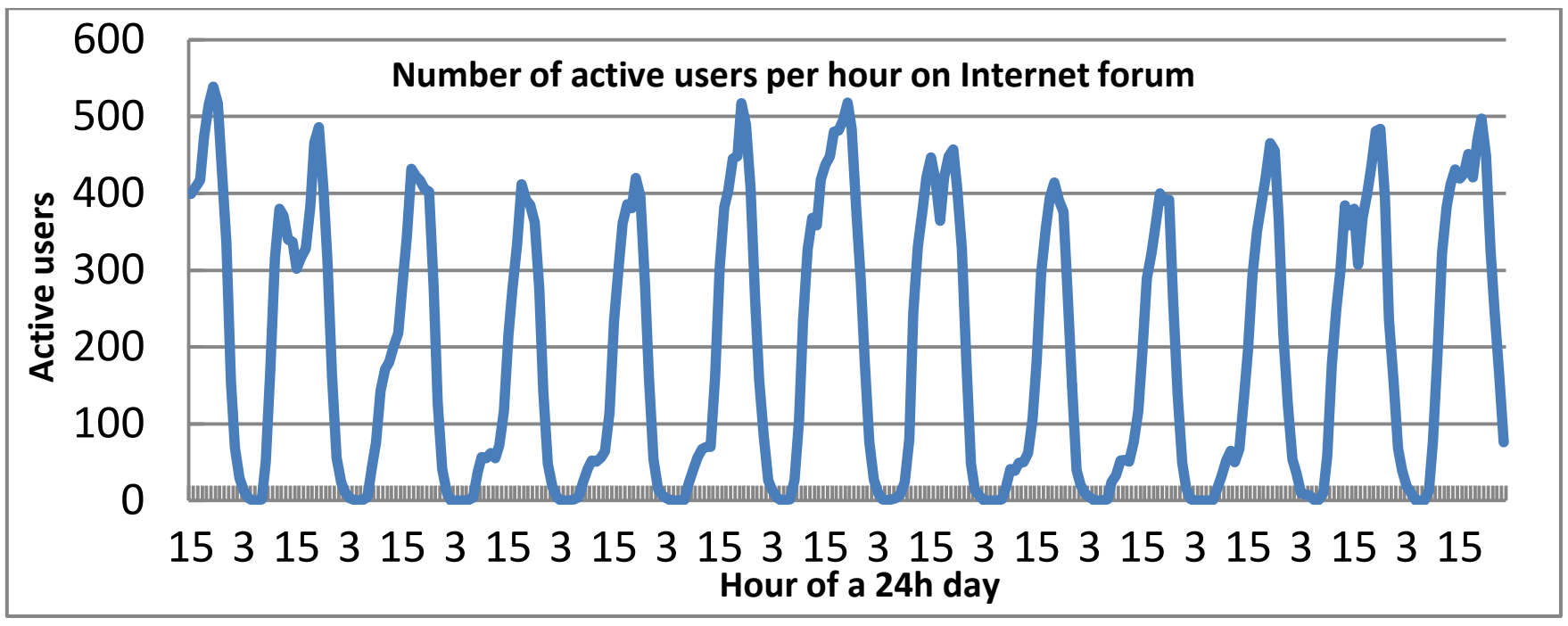

**Fig. 1.** Users activity during 13 days of two 'artificial' campaigns. Active user is defined as a one who sent/received at least one message or equivalent

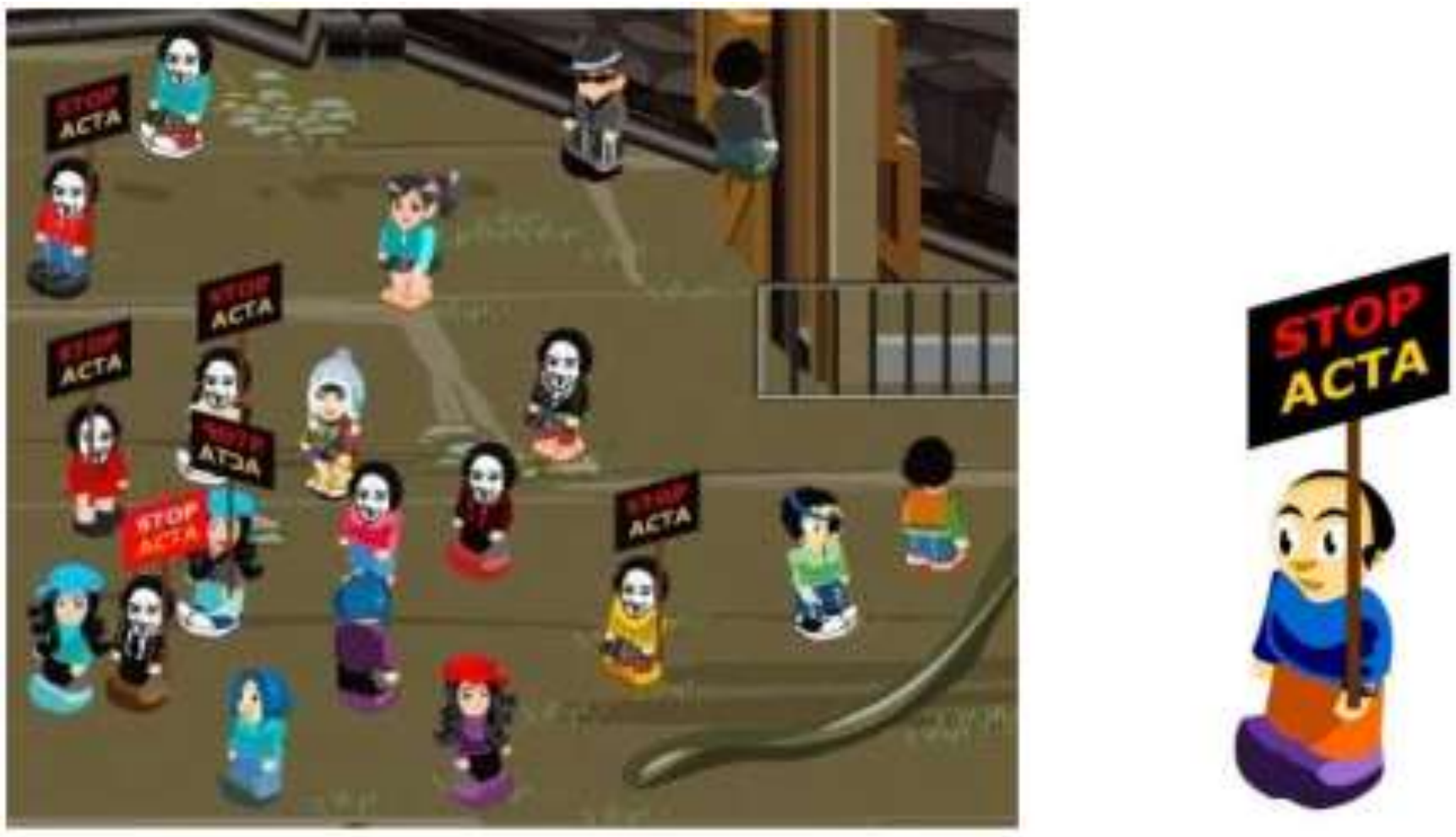

**Pic. 1.** Visualization of different avatars (right) and Stop-ACTA stand up (left)

2 After Poland's announcement that it will ratified the ACTA treaty, protests were held against, mainly conducted by Internet – users like our studied community

## 3 Study design

Epidemic spread can take place by external source of contact between individuals, who form different kind of social networks. Empirical research related to the question of factors driving social contagion come from diverse studies, but mostly based on networks. Nowadays, based on data gathered in computer systems, a new type of social network can be extracted and analyzed [12]. These networks are usually automatically extracted from such data sources as: bibliographic data, blogs, e-mail systems, telecommunication data, and social services like Twitter or Facebook [13], video sharing systems like YouTube [14], Wikipedia and many more. In our case individual could be infected by another individual. Spread of a viral can be described in terms of the epidemiological process. Let's use epidemiological notation of SEI (susceptible – exposed – infective) concept [1]. In this study sending invitation is like transmitted pathogen from I to S and receiver changes his state to E. If agent, who was in state E, decides to send further invitation, he becomes to be in state I. Process is irreversible. Because infective agents could send invitation to any active users (in both possible states S, E, I), we can differentiate two kinds of infections: unique (when agent in state S get his first invitation) and non-unique.

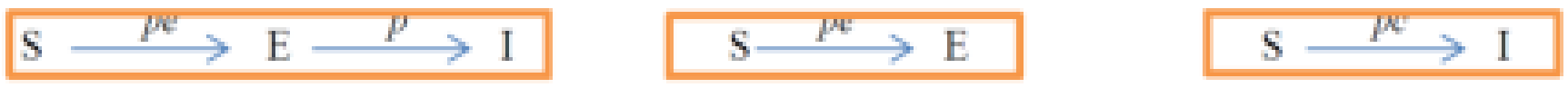

**Pic. 2.** Model definitions: full model (left), reduced model where transmission from E to I omitted (center), reduced model where state E is skipped (right)

## 4 Concept of data analyses

Periodicity, exchange of users [15] in virtual world makes observations more complicated [Fig. 2]. To cope with that, let us investigate the process only on two scales.

- First, global – to get general view and find general properties we observe whole campaign from stationary not time dependent perspective. We focus on epidemic global parameters. Unfortunately, we know, they are not stationary and changing with time or waves [16] (generations) [Fig. 2], but only this level comparisons between campaigns give opportunity to conclude something about difference in quantitative manner.
- Second, local –to get more insight in dynamic in time framework equal to the seasonality periods we observe first day of campaign. This approach gives opportunity to a reader to understand in more qualitative way what is happening in the system.

We choose only those small snapshots into account in this study, to expose role of emotions in viral spread, but in other papers about this virtual world [17] you can find other aspects from ego networks [18], [19] to predicting effects of marketing strategies [20], [21].

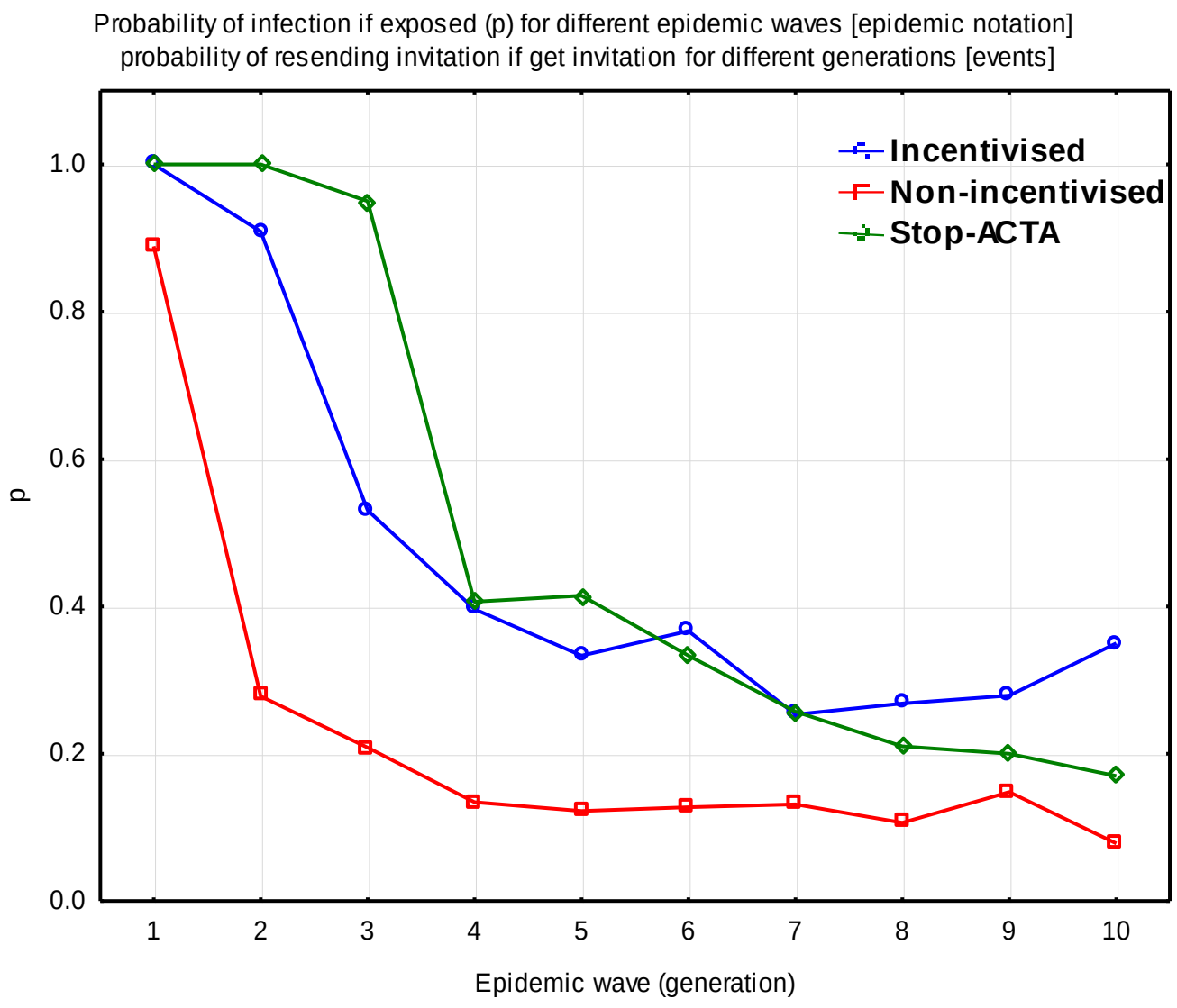

**Fig. 2.** Dynamics of campaign property - probability of infection if exposed (*p*) / probability of resending invitation if get invitation - for first 10 generations

# 5 Global analyses

Overall statistics of campaigns [Tab. 1] seems to be different, but we have only one realization of such each scenario, so random effect would play big role in terms of duration of epidemic of numbers of contacts or exposed or infective. More informative are model [Pic. 2] parameters [Tab. 2] which explain ruling behavior. The main difference for emotional study (anti-ACA) is observed for *pe* and *pc* values. Those probabilities are few times bigger for natural than both artificial experiments. Using epidemiological notation, probability to send infection for pair of individuals who were in contact (*pe*) is much bigger for Stop-ACTA studies. It can be understand, that viral were send in more precise way. Additionally, big value of probability of infection per contact (*pc*) means that such transmissions (invitation) are effective. Those two parameters tell us that emotional oriented information is targeted. In other worlds, we can claim that people are sending emotional information mainly to people, for whom that message can be relevant. On the other hand we have to add, that probabilities of infection if exposed (*p*) are similar (difference up to 50%), so probability of sending infection further do not depend as much on emotional aspect (standard deviation calculated based on aggregated generation data [Fig. 2] does not allow to claim any conclusions with statistical significance). As an example, we also can suggest, that probabilities of retwitting (in Twitter of Facebook) would not depend on emotional value of information (how it seems to be observed [22]).

| Event | all invitations | unique invi-tations | No. user who send invitation | No. of generations (until no more invita-tion is send) | Time of campaign |
|---|---|---|---|---|---|
| Epidemiological notation | Total No. Con-tacts | Total No. Exposed (E) | Total No. Infective (I) | No. of waves (until epidemic died out) | Time of epidemic |
| Non-incentivized | 9972 | 3069 | 746 | 12 | 13 days |
| **Stop-ACTA** | **731** | **635** | **242** | **14** | **44(5) days** |
| Incentivized | 28446 | 3874 | 1873 | 14 | 13 days |

**Table 1.** Campaign quantities

| Event | proportion of unique invitations | probability of resending invitation per invitation | probability of resending invitation if get invitation |
|---|---|---|---|
| Epidemiological notation | probability of being exposed per contact ($pe$) | probability of infection per contact ($pc$) | probability of infection if exposed ($p$) |
| Non-incentivized | 0.31 | 0.07 | 0.24 (std=0.23) |
| **Stop-ACTA** | **0.87** | **0.33** | **0.38 (std=0.28)** |
| Incentivized | 0.14 | 0.07 | 0.48 (std=0.26) |

**Table 2.**Model parameters calculated for campaigns

# 6 Local - dynamic analyses (first day of ant-ACTA campaign)

Trying to fit solution of standard time dependent SEI model, we found difficult to implement it directly because of user activity patterns [Fig. 1]. Such a behavior is difficult to capture, because standard dynamic model SEI require knowing total number of users. First, we reduced problem to SE model [Pic. 2 center]. This seems to be relevant for Stop-ACTA, because probability $pe$ is close to one [Tab. 2], moreover it is much higher than in artificial campaigns. Standard irreversible epidemic process, where receiving invitation is treated as an infection (SE) can be described by one differential equation[3]:

$$\frac{dE}{dt} = \frac{r}{N} SE \tag{1}$$

[3] In standard SE models (in literature called usually SI) whole population is infected, and marked simplification comes from this observation.

- $r$ is “transmission coefficient” per time unit, so it is time dependent equivalent of $pe$ [Pic. 2 center]. That definition assumes the same probability of infections as well as the same number of contacts for all users.
- $N(t) = S(t) + E(t) = const.$

Solution of stationary case ($r = const$) is given [23] by function with initial $E_0$ cases:

$$E_{total}(t) = E(t) = \frac{NE_0}{E_0 + (N - E_0)\exp(-rt)} \quad (2)$$

Main property of such a model tells that all individuals are infected ($E$) after some time. The same appears in our campaign that almost all users have got invitation. Unfortunately, we cannot use that model as it is because of constant exchange of users in time, combined with daily activity patterns. Instead of constant number of user let us allow to vary it. In all cases, total number of active users ($N$) for one day was implemented as a decreasing function from maximum amount of (around 350-450) users around afternoon to minimum (no users) during late nights hours. We used standard analytical solution of deterministic model of SE (2) and introduced time dependence to total number of active users ($N$) as an effective factor of infectivity ($r = r'N(t)$). The total number of infected is approximately described by:

$$E_{total}(t) = \frac{N_{inf}E_0}{E_0 + (N_{inf} - E_0)\exp(-r\prime N(t)t)} \quad (3)$$

Total number of expositions at end of the day ($E_{inf}$) plays a role of the finite asymptotic number of users in system ($N_{\mathrm{inf}}$). In more exact approach we introduced to standard model of SE (1) linear decrease (with coefficient $a$) of active users (4). It was implemented by imitation of decrease of total number of user from pick afternoon hours to no activity late at night [Fig. 1] by $N(t) = N_0 - at$ with respectively decrease of $E$ population.

$$\frac{dE}{dt} = \frac{r}{N_0 - at}(N_0 - at - E)E - a\frac{E}{N_0 - at} \quad (4)$$

Simulation of the same process took place to evaluate $E_{total}$. Fits of solutions (2), (3) and simulations results of the model (4) were showed with cumulative number of unique exposition in first 6 hours of Stop-ACTA campaign [Fig. 4]. Empirical process is too quick, in first phase comparing to analytical solutions, but slower than simulations. That can be explained by at least two factors. First, that infectivity is changing in time, what was really observed [Fig. 2]. Users are coming in or out and population is mixing, but we introduced only decrease of number of them [Fig. 1].

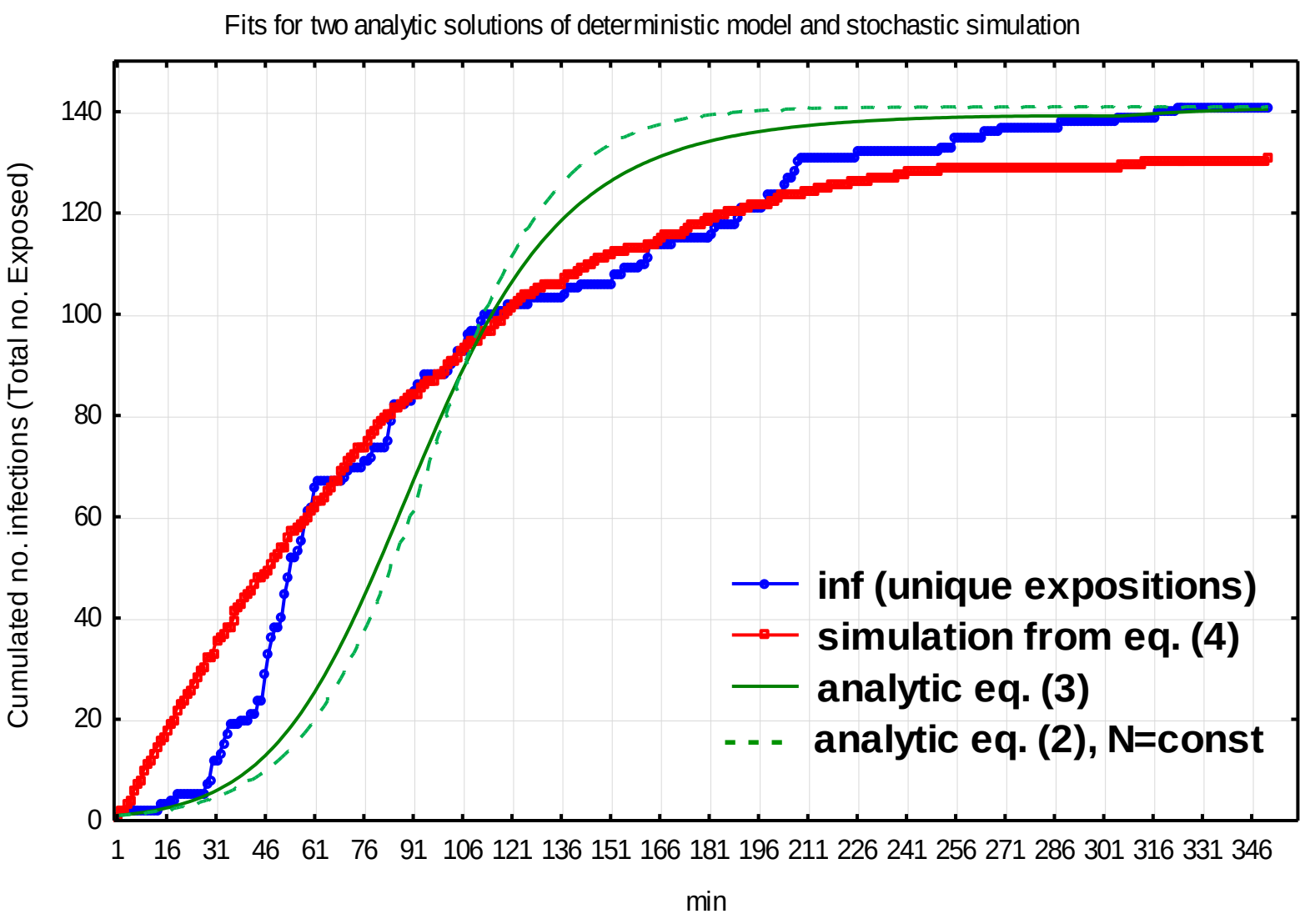

**Fig. 3.** Dynamics of Stop-ACTA campaign during fist day

The same evolution as in time-dependent case [Fig.3] can be observed in sequence generation representation [Fig.4]. Graph of 141 nodes (who get exposed - E in this short history) with 53 users who sent further infection is a tree with loops [24]. During first day of Stop-ACTA campaign 10 waves (generation) come to alive before process died out (only for a night, because viral was propagating since morning day after but local analyses stopped at first day only).

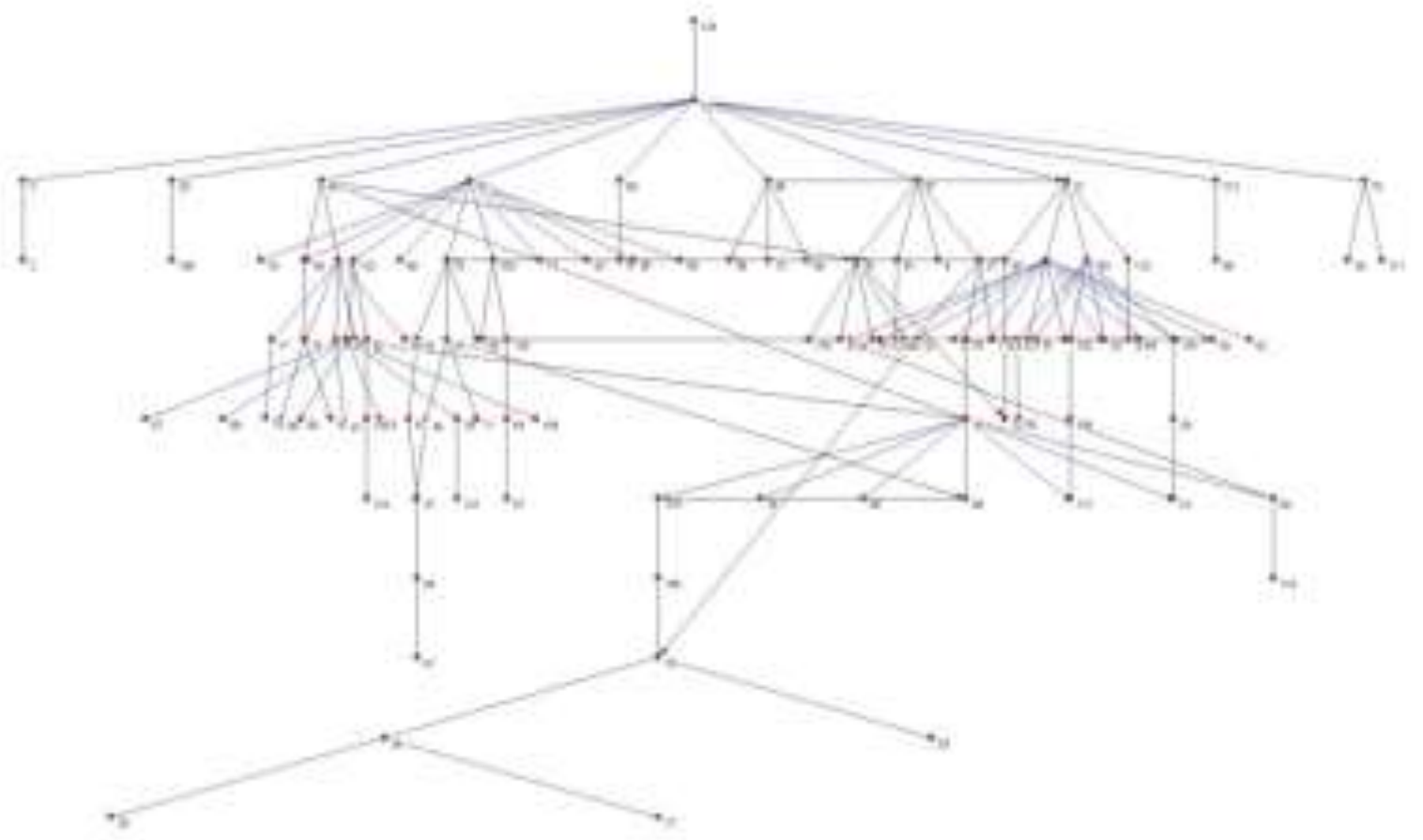

**Fig. 4.** Hierarchical representation of Stop-ACTA campaign during the first day. Links show branching process and arrows non-unique infections (invitations)

## 7 Conclusions and further work

The presented research showed epidemiological approach to viral campaign analysis by comparing emotional with artificial experiments. Social contagion studies are usually a simplification of a real world situation where decisions related to participation in viral diffusion are based on many different factors which are difficult to observe and monitor. Even than we observed that emotional messages [Tab. 2] are deliberately targeted (high *pe* and *p*c for Stop-ACTA) and less blindly forwarded (the same *p* for Stop-ACTA and both artificial). On the other hand, our conclusion are still below any statistical significance [Tab. 2] and external factors like media, mainly on Stop-ACTA campaign, cannot be controlled. By global analyses, it is possible to catch some deepen characteristics and interesting results, but local dynamic still cannot be described precisely [25] and needs further investigation [Fig. 3].The proposed model attempts to merge micro dynamics (spread of viral [Pic. 2]) with users behavior (users activities in virtual world [Fig. 1]). The study showed that model parameters are time or generation dependent [Fig.2]. Ongoing research will be focused on network-oriented agent based model of viral spread. Message exchange between users A and B, entries to individual profile, etc. will able us to constrain contact network, which was unknown in presented model. The properties of individuals, like frequency of expressing negative of positive emotion can be measured (e.g. by counting emoticons) and applied to model. A general study on hierarchical structure of viral spread [Fig. 4] is planned as well.

**Acknowledgment**

We would like to thank to people from Social Network Analyse Group at Wroclaw University of Technology and from Stockholm University for cooperation. AJ thanks to Swedish Institute for invitation to Sweden.

### Literature

[1] R. M. Anderson and R. M. May, Infectious Diseases of Humans: Dynamics and Control, Oxford: Oxford University Press, 1992.

[2] E. Rogers, Diffusion of Innovations, New York: Free Press, 2005.

[3] A. Nowak, J. Szamrej and B. Latané, "From private attitude to public opinion: A dynamic theory of social impact.," Psychological Review, vol. 97, pp. 362-376, 1990.

[4] D. Watts, "A simple model of global cascades on random networks.," PNAS, vol. 99, no. 9, pp. 5766-5771, 2002.

[5] A. Grabowski, "The relationship between human behavior and the process of epidemic spreading in a real social network," EPJ B, vol. 85, p. 248, 2012.

[6] F. Liljeros, "Information dynamics shape the sexual networks of Internet-mediated prostitution," PNAS, vol. 107, 2010.

[7] J. Leskovec, L. Adamic and B. Huberman, "The dynamics of viral marketing," ACM Trans. Web, vol. May, 2007.

[8] L. Rocha, Exporing patterns of empirical networks, PhD Thesis, Umea University, 2011.

[9] F. Liljeros, "The web of human sexual contacts," Nature, vol. 411, no. 6840, pp. 907-908, 2001.

[10] P. Holme, "Temporal networks," Physics Reports, vol. 519, no. 3, October 2012.

[11] A. Czaplicka and J. Hołyst, "Modeling of Internet Influence on Group Emotion," International Journal of Modern Physics C, vol. 23, p. 1250020, 2012.
[12] F. Liljeros, "Implementation of Web-Based Respondent-Driven Sampling among Men who Have Sex with Men in Vietnam," PLoS One, vol. 7(11), 2012.
[13] S. Goel, D. J. Watts and D. G. Goldstein, "The Structure of Online Diffusion Networks," ACM Journal, vol. February, 2012.
[14] Y. Liu-Thompkins, "Seeding Viral Content: Lessons from the Diffusion of Online Videos," Journal of Advertising Research, vol. Nov, 2011.
[15] A. Grabowski and R. Kosiński, "Life span in online communities," Phys. Rev. E, vol. 82, p. 066108, 2010.
[16] C. Jacob, "Branching Processes: Their Role in Epidemiology," International Journal of Environmental Research and Public Health, vol. 7, p. 1186–1204, 2010.
[17] J. Jankowski, "Analysis of Multiplayer Platform Users Activity Based on the Virtual and Real Time Dimension," LNCS, vol. 6984, pp. 312-315, 2011.
[18] J. Jankowski, S. Ciuberek, A. Zbieg and R. Michalski, "Studying Paths of Participation in Viral Diffusion Process," LNCS, no. 7710, pp. 503-516, 2012.
[19] A. Zbieg, B. Żak, J. Jankowski, R. Michalski and S. Ciuberek, " Studying Diffusion of Viral Content at Dyadic Level," in ASONAM 2012, Istambul, 2012.
[20] J. Jankowski, R. Michalski and P. Kazienko, "The Multidimensional Study of Viral Campaigns as Branching Processes," LNCS, vol. 7710, pp. 462-474, 2012.
[21] R. Michalski, J. Jankowski and P. Kazienko, "Negative Effects of Incentivised Viral Campaigns for Activity in Social Networks," in SCA, China, 2012.
[22] M. Thelwall, K. Buckley and G. Paltoglou, "Sentiment strength detection for the social Web", Journal of the American Society for Information Science and Technology, 63(1), pp. 163-173, 2012.
[23] A. Jarynowski, "Human-human interaction: epidemiology," in Life time of correlation, Wrocław, Wydawnicwto Niezależne, 2010.
[24] Z. Burda, J. Correia and K. A., "Statistical ensemble of scale-free random graphs," Physical Review E, vol. 64, pp. 461181-461189, 2001.
[25] K. Malarz, Z. Szvetelszky, B. Szekfu and K. Kułakowski, "Gossip in random networks," Acta Physica Polonica B, vol. 37, no. 11, pp. 3049-3058, 2006.